\documentclass{article}
\usepackage{spconf,amsmath, amsfonts,graphicx}
\usepackage{xcolor}

\usepackage{caption}
\usepackage{subcaption}
\usepackage{hyperref}
\usepackage{booktabs}
\usepackage{multirow}

\usepackage{enumitem}
\setlist[itemize]{noitemsep, nolistsep}
\usepackage{fancyhdr}

\newcommand{\etal}{\textit{et~al}.}

\newcommand{\eg}{\textit{e}.\textit{g}., }

\DeclareMathOperator{\softmax}{softmax}

\newcommand{\method}{GABIC}

\title{\method: Graph-based Attention Block for Image Compression}
\name{
\begin{tabular}{c}
Gabriele Spadaro$^{1,2}$ \quad Alberto Presta$^1$ \quad Enzo Tartaglione$^2$ \quad Jhony H. Giraldo$^2$ \\
Marco Grangetto$^1$ \quad Attilio Fiandrotti$^{1,2}$
\end{tabular}
}

\address{$^1$ University of Turin, Italy\\
$^2$ LTCI, T\'el\'ecom Paris, Institut Polytechnique de Paris, France}

\fancypagestyle{first_page_style}
{
   \fancyhf{}
   \lfoot{\color{gray}\scriptsize © 2024 IEEE. Personal use of this material is permitted. Permission from IEEE must be obtained for all other uses, in any current or future media, including reprinting/republishing this material for advertising or promotional purposes, creating new collective works, for resale or redistribution to servers or lists, or reuse of any copyrighted component of this work in other works.}
}

\begin{document}
%
\maketitle

\begin{abstract}


While standardized codecs like JPEG and HEVC-intra represent the industry standard in image compression, neural Learned Image Compression (LIC) codecs represent a promising alternative.
In detail, integrating attention mechanisms from Vision Transformers into LIC models has shown improved compression efficiency.
However, extra efficiency often comes at the cost of aggregating redundant features.
This work proposes a Graph-based Attention Block for Image Compression (\method), a method to reduce feature redundancy based on a $k$-Nearest Neighbors enhanced attention mechanism. 
Our experiments show that \method~outperforms comparable methods, particularly at high bit rates, enhancing compression performance.
\end{abstract}


\fancyhf{}
\renewcommand{\headrulewidth}{0pt}
\setcounter{page}{1}
\cfoot{\thepage}
\thispagestyle{first_page_style}
\section{Introduction}

Standardized image codecs like JPEG~\cite{jpeg}, the HEVC-based BPG~\cite{bpg}, and VVC~\cite{vvc} rely on a clever combination of coding tools such as transform coding, quantization, and entropy coding to achieve state-of-the-art in terms of Rate-Distortion (RD) efficiency.
Recently Learned Image Compression (LIC) based on deep neural architectures such as Variational Autoencoders (VAEs) have gained traction in reason of their
encoding efficiency.
After the seminal work~\cite{balle17}, most LIC architectures have predominantly used Convolutional Neural Networks (CNNs)~\cite{balle2018variational,minnen2018joint, cheng2020, zou2022devil} improved with Generalized Divisive Normalization (GDN) layers~\cite{gdn}, within a VAE framework.

\begin{figure}[t]
    \centering
    \includegraphics[width=\columnwidth]{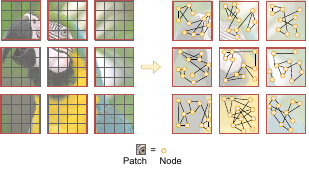}
    \caption{Dividing the image into patches, it is possible to compress their content using a GNN. This allows better preserving of local details due to the dynamic graph structure on which the attention is computed, enabling higher compressibility at high bit rates.}
    \label{fig:teaser}
\end{figure}

Recently, attention mechanisms from Vision Transformers (ViTs) have been incorporated into LIC models~\cite{cheng2020,zou2022devil,liu2023tcm} showing improved compression efficiency~\cite{zou2022devil}.
However, the windowing-based self-attention mechanism used, \eg in Swin Transformers~\cite{swin}, may result in redundant features that add to the rate of the compressed latent representation.
We point out that the self-attention mechanism in ViTs can be seen as a graph attention network~\cite{gat} operating within a locally fully connected graph.
The intuition behind this work is that if the self-attention mechanism could be locally constrained within a local graph, that may avoid the aggregation of redundant visual features,
as illustrated in Fig.~\ref{fig:teaser}.
Understanding how to efficiently constrain the attention mechanism is a challenging problem that, to the best of our knowledge, has not been tackled yet motivating the present work.

In this paper, we
introduce a novel attention mechanism called \textbf{G}raph-based \textbf{A}ttention \textbf{B}lock for \textbf{I}mage \textbf{C}ompression (\method). 
In a nutshell, \method~relies on a graph-based $k$-Nearest Neighbors ($k$-NN) mechanism at attention time to reduce the number of redundant features.
The contributions of this work can be summarized as follows:
\begin{itemize}
    \item To the best of our knowledge, \method~is the first local graph-based attention mechanism that allows clustering redundant features (Sec.~\ref{ch:graph_win_block}) employed in an end-to-end compression model.
    \item Our experimental results show that our method improves the encoding efficiency with respect to comparable approaches, especially at lower distortion (Sec.~\ref{sec:rdcomparison}). 
    \item We publicly released both the code and the trained models at \href{https://github.com/EIDOSLAB/GABIC}{https://github.com/EIDOSLAB/GABIC}.
\end{itemize}

\section{Related Work}
\label{sec:related_work}

This section provides some background on LIC and reviews the relevant literature (Sec.~\ref{lic_back}), with a main focus on both how attention mechanisms have been exploited and the potentiality of graph-based modules (Sec.~\ref{sec:attlic}).
 
\subsection{Learned Image Compression} 
\label{lic_back}

In LIC, an image is compressed employing a parametric autoencoder, trained end-to-end using backpropagation~\cite{review}.
In particular, an encoder $f_{a}(\cdot)$ projects an image $\mathbf{x}$ to a latent representation $\mathbf{y}$, that is first quantized into $\mathbf{\hat{y}}$ and then entropy-coded as a compressed bitstream.
On the receiving side, the latter is first retrieved and then projected back to its original domain by a decoder $f_{s}(\cdot)$, which outputs the reconstruction of the image $\mathbf{\hat{x}}$.
The entire framework is trained by optimizing the well-known Rate-Distortion loss function:
\begin{equation}
    \mathcal{L} = R(\mathbf{\hat{y}}) + \lambda D(\mathbf{x}, \mathbf{\hat{x}}),
    \label{eqn:RD_loss_function}
\end{equation}
where $\lambda$ is the hyper-parameter that controls the trade-off between the rate $R(\cdot)$ and the distortion $D(\cdot)$, that can be any distance metric between input $\mathbf{x}$ and reconstruction $\mathbf{\hat{x}}$ (like a mean squared error), and $R(\cdot)$, representing the estimated rate of $\mathbf{\hat{y}}$.
The loss in \eqref{eqn:RD_loss_function} accounts for both quality of the reconstruction $\mathbf{\hat{x}}$ and compressibility of the latent representation $\mathbf{\hat{y}}$.

The first studies like~\cite{balle17} exploited a simple structure with only one latent representation, modeled as a fully factorized distribution extracted either with an auxiliary neural network or in an analytical way~\cite{presta23}.
More recent architectures rely on a further representation $\mathbf{\hat{z}}$, called \emph{hyperprior}~\cite{balle2018variational}, extracted from a second encoder-decoder pair $\{h_a(\cdot),h_s(\cdot)\}$ to capture spatial correlations from $\mathbf{y}$. 
In this scenario, $\mathbf{\hat{y}}$ and $\mathbf{\hat{z}}$ are modeled with a Gaussian and channel-wise fully-factorized distribution, respectively. 

Various works have been developed upon this architecture, to increase the overall performance.
For example, Minnen~\etal~\cite{minnen2018joint}~introduced an autoregressive model based on mask convolution to extract more context information.
Minnen and Singh~\cite{minnen2020channel}~developed a channel-wise entropy estimation model that first divides the latent representation into slices and then encodes them sequentially.
Other works tried to exploit other techniques, \eg Xie~\etal~\cite{xie2021}~introduced invertible modules to preserve information through layers, while Yang~\etal~\cite{yang2021graph} exploited graph-convolution to extract both local and global features at the same time.

\subsection{Attention Mechanism in LIC}
\label{sec:attlic}

Motivated by the success of attention mechanisms in tasks related to natural language processing and computer vision, many tried to introduce attention modules in the compression framework to optimize the bitrate allocation.
To capture long-range dependencies between pixels, non-local attention blocks were first integrated within the end-to-end image compression frameworks \cite{cheng2020,Zhou2019e2eAtt,Liu2019NonlocalAO}.
In \cite{zou2022devil}, the authors notice instead that generating attention maps based on spatially neighboring
elements can improve RD performance with fewer computations.
For this reason, they replaced the non-local attention module of~\cite{Liu2019NonlocalAO} with a window block for focusing on regions with high contrast and to put more effort into image details.
In a similar spirit, \cite{liu2023tcm}~proposed a parallel Transformer-CNN mixture block with a controllable complexity, able to join the local modeling ability of the convolution and the non-local modeling ability of transformers.
Besides, they introduced a Swin-attention module in the channel-wise entropy estimation architecture.

\subsection{Vision Graph Neural Networks}

Meanwhile, Graph Neural Networks (GNNs), typically used for graph-based data~\cite{kipf2017semi}, have demonstrated remarkable success when applied to tasks such as image classification~\cite{vig, hvig} and segmentation~\cite{giraldo2022hypergraph}.
Drawing inspiration from the partition concept introduced in ViTs~\cite{vit}, the input image undergoes segmentation into smaller patches, with each patch representing a node in the graph of the image. 
Connections between nodes are established using the $k$-NN technique in the feature space.
Subsequently, graph convolution operators are employed to update the features of each node, considering its position in the graph~\cite{gilmer2017neural}.
Following a similar paradigm as Transformers, these features contribute to the classification of the entire graph, thereby classifying the entire image.

Several works showed that the attention mechanism is key for LIC models' compression efficiency.
For example, the work by Liu~\etal~\cite{Liu2019NonlocalAO} introduced the Non-Local Attention Module (NLAM), consisting of a cascade involving a non-local block and regular convolutional layers.
Subsequently, Zhu~\etal~\cite{zou2022devil} replaced the non-local block with a window block to capture local dependencies. 
However, none of the above approaches exploits the full potential of GNNs that, to the present date, remain largely untapped.
Our proposed method, \method, marks a significant step forward in this direction, representing the inaugural endeavor to exploit the power of attention mechanisms coupled with GNNs for image compression.

\section{Method}
\label{sec:method}

\begin{figure*}
     \centering
     \begin{subfigure}[b]{0.74\textwidth}
         \centering
         \includegraphics[width=\textwidth]{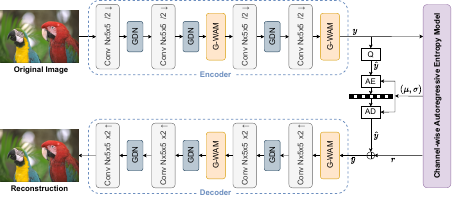}
         \caption{~}
         \label{fig:arc}
     \end{subfigure}
     \hfill
     \begin{subfigure}[b]{0.25\textwidth}
         \centering
         \includegraphics[width=.75\textwidth]{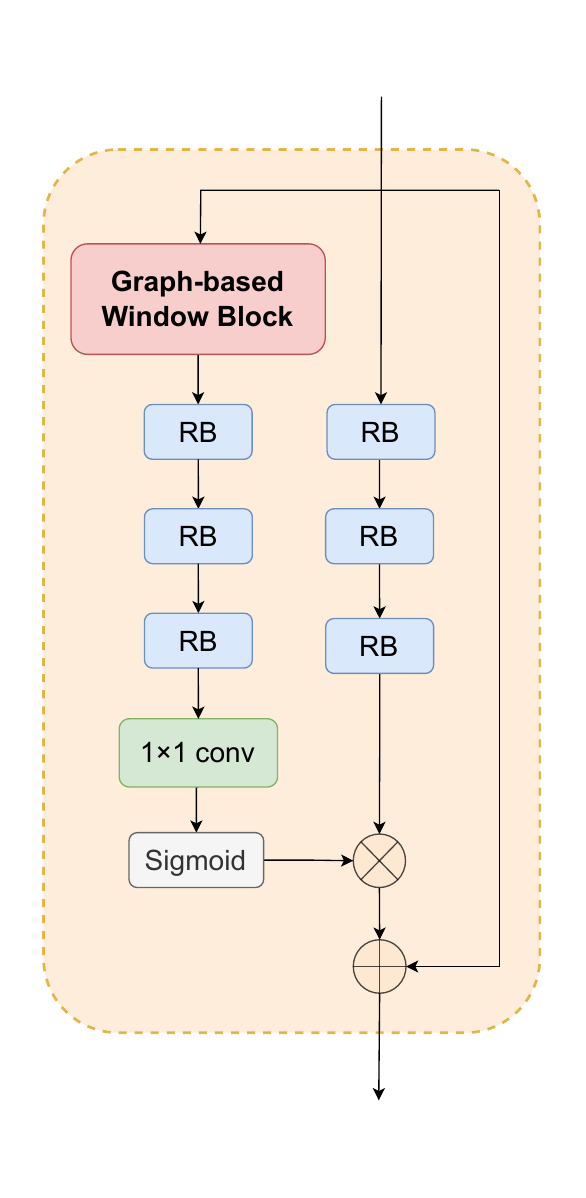}
         \caption{~}
         \label{fig:gwam}
     \end{subfigure}
     \hfill
    \caption{Overview on \method's architecture (a) and detail of the proposed Graph-based Window Attention Module (b).}
    \label{fig:architecture}
\end{figure*}

Fig.~\ref{fig:architecture} shows the general architecture of a state-of-the-art LIC model enhanced with our attention mechanism \method.
The model comprises an encoder-decoder pair, with attention computation facilitated by our graph-based window block within a graph structure. 
Additionally, the architecture builds upon the hyper-prior architecture \cite{balle2018variational} and incorporates a channel-wise entropy model \cite{minnen2020channel}, detailed in Sec.~\ref{sec:hyp}.
Fig.~\ref{fig:gwam} illustrates our Graph Window Attention Module (G-WAM), while Fig.~\ref{fig:arc} showcases the integration of these modules into both encoder and decoder pairs of our architecture.

\subsection{Preliminaries}

Let us represent a graph as the mathematical entity $~{G=(\mathcal{V},\mathcal{E})}$, where $~{\mathcal{V}=\{1,\dots,N\}}$ is the set of $N$ nodes and ${\mathcal{E}\subseteq \{(i,j)\mid i,j\in \mathcal{V}\;{\textrm {and}}\;i\neq j\}}$ is the set of edges between nodes $i$ and $j$.
In GNNs, we typically update each node embedding $i$ using a message-passing mechanism within the set of neighbors $\mathcal{N}(i)$ of node $i$ \cite{gilmer2017neural}.
\method~also uses the message-passing operations coupled with attention.

\subsection{Architecture}

\label{sec:hyp}
\begin{figure*}
     \centering
     \begin{subfigure}[b]{0.45\textwidth}
         \centering
         \includegraphics[width=\textwidth]{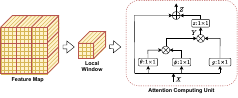}
         \caption{~}
         \label{fig:local_att}
     \end{subfigure}
     \hfill
     \begin{subfigure}[b]{0.45\textwidth}
        \centering
         \includegraphics[width=\textwidth]{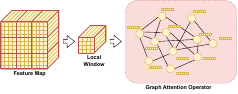}
         \caption{~}
         \label{fig:graph_att}
        
     \end{subfigure}
    \caption{Traditional local window block scheme (a) and our proposed Local Graph-based Window block (b).}
    \label{fig:three graphs}
\end{figure*}
\noindent
The above LIC model is characterized by a hyperprior-based architecture \cite{balle2018variational} and a channel-wise entropy model \cite{minnen2020channel} as shown in Fig.~\ref{fig:arc}.
Thus, the model can be formulated as:
\begin{equation} \label{gen_eq}
    \begin{split}
        \mathbf{y} &= f_{a}(\mathbf{x};\boldsymbol{\phi})\\
        \mathbf{\hat{y}} &= Q(\mathbf{y} - \mathbf{\mu}) + \mathbf{\mu} \\
        \mathbf{\hat{x}} &= f_{s}(\mathbf{\hat{y}};\boldsymbol{\theta}),
    \end{split}
\end{equation}
where $\boldsymbol{\phi}$ and $\boldsymbol{\theta}$ are trainable parameters of the encoder $f_{a}(\cdot)$ and the decoder $f_{s}(\cdot)$ respectively, while  $Q(\cdot)$ is the rounding quantization function.  
As mentioned in Sec.~\ref{lic_back}, $\mathbf{y}$ is estimated to have Gaussian distribution with mean $\mathbf{\mu}$. 
Following previous works \cite{minnen2018joint, zou2022devil}, we round and encode  $\lceil \mathbf{y} - \mathbf{\mu} \rfloor$ instead of $\lceil \mathbf{y} \rfloor$, as it has been proven to benefit the entropy models; we exploit a range coder to loosely encode  $\lceil \mathbf{y} - \mathbf{\mu} \rfloor$, which is modeled with a Gaussian distribution with variance $\mathbf{\sigma}$, and it is transmitted to the decoder $f_{s}(\cdot)$.
We extract $\mathbf{\mu}$ and $\sigma$ by leveraging both the hyperprior encoder-decoder  $h_a(\cdot),h_s(\cdot)$ and the channel-wise entropy model.
In particular, $\mathbf{y}$ is divided into $s$ slices which are then encoded sequentially so that the already encoded slices can improve the encoding of the current one;
furthermore, a slice network $\varphi_{i}(\cdot)$ has been introduced to reduce the error due to quantization artifact.
The overall pipeline can be formulated as follows:
\begin{equation} \label{entropy_eq}
    \begin{split}
        \mathbf{z} = h_{a} (\mathbf{y}, \phi_{h}), & \quad \mathbf{\hat{z}} = Q(\mathbf{z}) \\
        (\mathcal{D}_{\mu},\mathcal{D}_{\sigma}) &= h_{s}( \mathbf{\hat{z}}, \theta_{h}) \\
        \mathbf{r}_{i}, \mathbf{\mu}_{i},\mathbf{\sigma}_{i} &= \varphi_{i}(\mathcal{D}_{\mu},\mathcal{D}_{\sigma}, \mathbf{\bar{y}}_{<i},\mathbf{y}_{i}) \\
        \mathbf{\hat{y}}_i  &=  Q(\mathbf{y}_{i}  - \mathbf{\mu}_i) + \mathbf{\mu}_{i} \\
        \mathbf{\bar{y}}_{i} &= \mathbf{r}_{i} + \mathbf{\hat{y}}_{i},
    \end{split}
\end{equation}
where $0 \le i < s$ and  $\mathbf{\bar{y}}_{<i} = \{\mathbf{\bar{y}}_{0}, \mathbf{\bar{y}}_{1}, \dots,\mathbf{\bar{y}}_{i-1}\}$ represents the slices already encoded.
As presented in \eqref{entropy_eq}, the hyperprior autoencoder is meant to capture spatial dependencies among $\mathbf{y}$ employing $(\mathcal{D}_{\mu},\mathcal{D}_{\sigma})$; for a specific slice, the latter are used in combination with  $\mathbf{\bar{y}}_{<i}$ and $\mathbf{y}_{i}$ as input to the slice network $\varphi_{i}(\cdot)$, yielding both the Gaussian parameters and a residual vector $\mathbf{r}_{i}$, which has been introduced to reduce the quantization error. 
Once passed toward all the slices, $\mathbf{\bar{y}}$ is entered in the decoder $f_{s}(\cdot)$ instead of $\mathbf{\hat{y}}$ in \eqref{gen_eq}, obtaining the reconstructed image $\mathbf{\hat{x}}$.

The model is finally trained minimizing a rate-distortion loss function formulated as: 
\begin{equation}
    \begin{split}
        \mathcal{L} 
        = \mathbb{E}_{\mathbf{x}\sim p_{\mathbf{x}}}[-\log_{2} p_{\mathbf{\hat{y}}|\mathbf{\hat{z}}}({\mathbf{\hat{y}}|\mathbf{\hat{z}}}) &- \log_{2}p_{\mathbf{\hat{z}}} (\mathbf{\hat{z}})] \\
        &   + \lambda \mathbb{E}_{\mathbf{x}\sim p_{\mathbf{x}}}[d(\mathbf{x}, \mathbf{\hat{x}})].
    \end{split}
\end{equation}

\subsection{Graph Window Attention Block}
\label{ch:graph_win_block}

\method~computes local attention maps over non-overlapping windows.
However, we do not compute the self-attention mechanism as in regular Swin Transformer models \cite{zou2022devil,swin}, but instead we treat each patch of these windows as a node of a graph.
Therefore, we use the graph to compute the attention operation.
Fig.~\ref{fig:graph_att} shows an illustration of our updating operation (in contrast to the conventional self-attention in Fig.~\ref{fig:local_att}), where the input feature map is divided into $M \times M$ windows.
Let $\mathbf{x}(i)$ be the embedding of the patch $i$ in the local window.
We update the embedding of the patch $i$ using a graph convolution operation described as follows:
\begin{equation}
    \mathbf{x}_{\text{upd}}(i) = \mathbf{x}(i) + \mathbf{W}_{z} \sum_{j \in \mathcal{N}(i)} \alpha_{i,j} \mathbf{W}_{g}\mathbf{x}(j),
    \label{eqn:graph_convolution}
\end{equation}
where $\mathbf{W}_{z}$ and $\mathbf{W}_{g}$ are learnable parameters for the specific window, and $\alpha_{i,j}$ are the attention coefficients computed as:
\begin{equation}
    \begin{aligned}
        \alpha_{i,j} &= \softmax \left[(\mathbf{W}_{\theta}\mathbf{x}(i))^{\top} (\mathbf{W}_{\phi}\mathbf{x}(j))\right]
        \\
        &= \frac{\exp\left[(\mathbf{W}_{\theta}\mathbf{x}(i))^{\top} (\mathbf{W}_{\phi}\mathbf{x}(j))\right]}
        {\sum_{n \in \mathcal{N}(i)} \exp\left[(\mathbf{W}_{\theta}\mathbf{x}(i))^{\top} \mathbf{W}_{\phi}\mathbf{x}(n)\right]}.
    \end{aligned}
    \label{eqn:attention_coefficients}
\end{equation}
We dynamically update the neighbors of $i$ on each iteration using a $k$-NN in the space of features such that:
\begin{equation}
    \mathcal{N}(i) = \left\{ (i,j) : j \in k\text{-NN}\left[\mathbf{x}(i)\right] \right\},
\end{equation}
where $k\text{-NN}[\mathbf{x}(i)]$ is the set of $k$-NN of the feature $\mathbf{x}(i)$.\\

\noindent \textbf{Relationship with the conventional attention operation.}
We can recover the standard attention operation of the Swin Transformer~\cite{swin} in \eqref{eqn:graph_convolution} if $k$ is equal to the elements in the local window.
In other words, if $\mathcal{E}= \mathcal{V} \times \mathcal{V}$ in the local graph we recover the model proposed in \cite{zou2022devil} (illustrated in Fig.~\ref{fig:local_att}).
Therefore, the local attention operation can be defined as:
\begin{equation}
    \mathbf{y}(i) = \frac{1}{C(\mathbf{X})} \sum_{j \in \mathcal{V}} f\left[\mathbf{x}(i), \mathbf{x}(j)\right] g\left[\mathbf{x}(j)\right],
    \label{eq:local_att}
\end{equation}
and applying a residual connection for better convergence \cite{nonlocal_nn} we get:
\begin{equation}
    \mathbf{x}_{\text{upd}}(i) = \mathbf{x}(i) +\mathbf{W}_{z}\mathbf{y}(i).
    \label{eq:res_local_att}
\end{equation}
Here $f(\cdot, \cdot)$ is an extension of the Gaussian function used to compute the similarity in the embedding space, while $C(\cdot)$ is a normalizing factor defined as 
$C(\mathbf{X}) = \sum_{j \in \mathcal{V}} f[\mathbf{x}(i), \mathbf{x}(j)]$, generally implemented as:
\begin{equation}
    \begin{aligned}
    f[\mathbf{x}(i), \mathbf{x}(j)] &= \exp{\left\{\theta\left[\mathbf{x}(i)\right]^{\top} \phi\left[\mathbf{x}(j)\right]\right\}}\\[.5em]
    \theta[\mathbf{x}(i)] &= \mathbf{W}_{\theta} \mathbf{x}(i) \\
    \phi[\mathbf{x}(j)] &= \mathbf{W}_{\phi} \mathbf{x}(j),
    \end{aligned}
\end{equation}
and using $g[\mathbf{x}(j)] = \mathbf{W}_{g} \mathbf{x}(j)$.
However, as notice by Wang~\etal~\cite{nonlocal_nn}, for a given $i$, $\frac{1}{C(\mathbf{X})} f[\mathbf{x}(i), \mathbf{x}(j)]$ becomes the $\softmax$ computation along the dimension $j$.
Thus, \eqref{eq:local_att} can be rewritten as:
\begin{equation}
    \mathbf{y}(i) = \sum_{j \in \mathcal{V}} \softmax \left\{\theta[\mathbf{x}(i)]^{\top} \phi[\mathbf{x}(j)] \right\} g(\mathbf{x}(j)),
\end{equation}
and consequently \eqref{eq:res_local_att} can be reformulated as follows:
\begin{equation}
\begin{aligned}
    \mathbf{x}_{\text{upd}}(i) &= \mathbf{x}(i) + \mathbf{W}_{z} \sum_{j \in \mathcal{V}} \alpha_{i,j} \mathbf{W}_{g}\mathbf{x}(j) \\
    \alpha_{i,j} &= \softmax \left\{[\mathbf{W}_{\theta}\mathbf{x}(i)]^{\top} \mathbf{W}_{\phi}\mathbf{x}(j)\right\}.
\end{aligned}
\end{equation}
With this last formulation, it is easy to spot the analogy with our adopted graph convolutional operator reported in~\eqref{eqn:attention_coefficients}.

\begin{figure*}[!ht]
     \centering
     \begin{subfigure}[b]{0.49\textwidth}
         \centering
         \includegraphics[width=.8\textwidth]{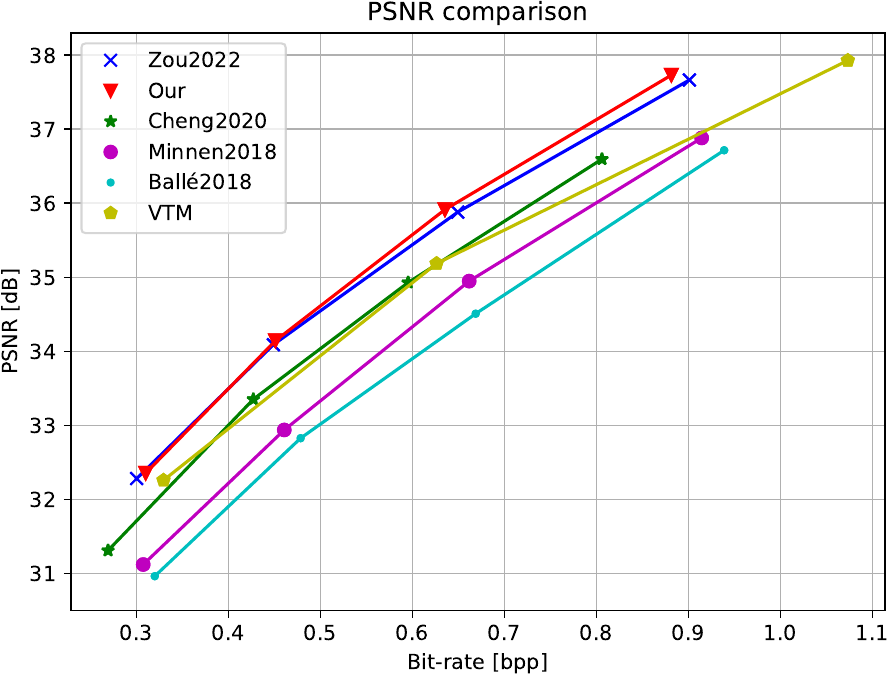}
         \caption{Experiments on the Kodak dataset.}
         \label{fig:rd_kodak}
     \end{subfigure}
     \hfill
     \begin{subfigure}[b]{0.49\textwidth}
         \centering
         \includegraphics[width=.8\textwidth]{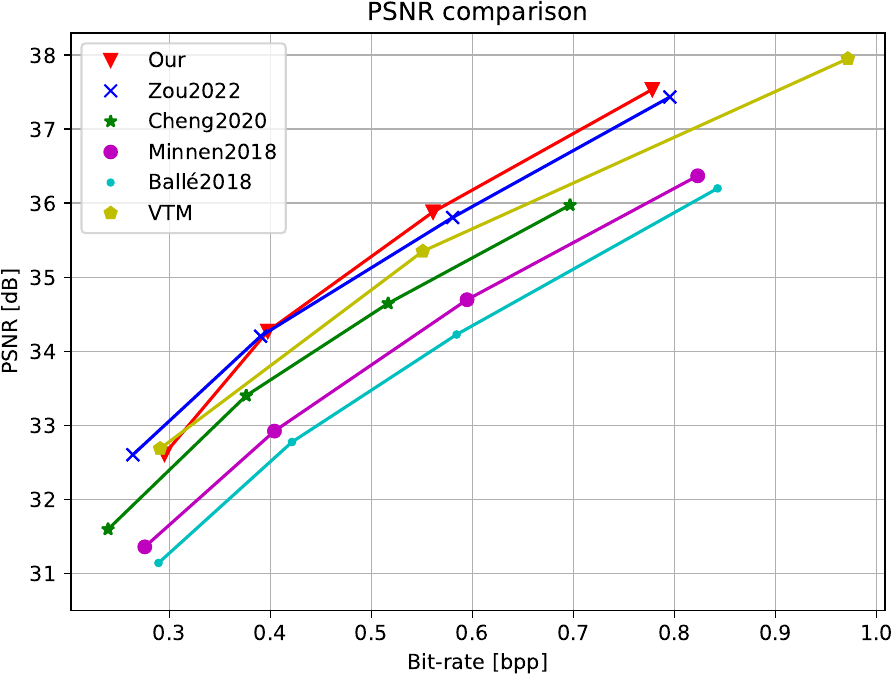}
         \caption{Experiments on the CLIC dataset.}
         \label{fig:rd_clic}
     \end{subfigure}
     \hfill
    \caption{Rate Distortion plots for the Kodak (a) and CLIC (b) datasets.}
    \label{fig:rd_res}
\end{figure*}

\section{Experiments}
\label{sec:experiments_results}

\subsection{Experimental Setup}
\noindent \textbf{Training.} To facilitate comparisons with other end-to-end compression models, our proposed graph-based architecture was implemented in the CompressAI platform~\cite{begaint2020compressai}, and the local Graph-based Window Block in Pytorch-Geometric~\cite{pyg}. The training was conducted by randomly selecting $300$k images from the OpenImages dataset~\cite{openimages} and randomly cropping them with the size of $256 \times 256$.
We trained our models for $200$ epochs using Adam as optimizer~\cite{kingma2015adam}, a batch size of $16$, and an initial learning rate equal to $1\times 10^{-4}$ and then decremented by a factor of $0.3$ with patience of $10$ epochs. Following the setting reported in \cite{begaint2020compressai}, we optimized our models using the \emph{Mean Squared Error} (MSE) as distortion term in the loss, and using lambda values $\lambda$ belonging to $\{0.0067, 0.0130, 0.025, 0.0483\}$.

As done by Zou~\etal~\cite{zou2022devil}, we configured the number of channels 
for the latent and hyper-latent spaces to $320$ and $192$, respectively. Additionally, we set the window size for the first and second Graph-based Window Blocks in both the encoder and decoder to $8$ and $4$, respectively.\\

\noindent\textbf{Evaluation.}
We evaluate \method~in terms of RD performance (PSNR) on the commonly used Kodak image set \cite{kodak} and the first $30$ images of the CLIC professional dataset~\cite{clic}. 
We compare our method with the VCC-VTM \cite{vvc} standard codec and other well-known learned compression models, including the context-free hyperprior model (Ball\'{e}2018)~\cite{balle2018variational}, auto-regressive hyper-prior model (Minnen2018)~\cite{minnen2018joint}, and auto-regressive hyper-prior model with GMM and simplified attention (Cheng2020)~\cite{cheng2020}. We also compare our results with Zou2022~\cite{zou2022devil}, in which the standard local attention is implemented. In Tab.~\ref{tab:compl} we show that our graph-based module preserves the same complexity as the local-attention one.

\begin{table}[!t]
\centering
\resizebox{0.45\textwidth}{!}{
\begin{tabular}{@{}lcccc@{}}
\toprule
Method & Enc (ms) & Dec (ms) & FLOPs ($10^{9}$) & Memory (Mb)\\
\midrule
Minnen2018~\cite{minnen2018joint} & $65.7_{\pm 1.5}$ & $54.1_{\pm 1.3}$ & $94.32$ & $72.37$\\
Cheng2020~\cite{cheng2020} & $3194.9_{\pm 8.2}$ & $8844.6_{\pm 16.5}$ & $831.35$ & $126.95$\\
Zou2022~\cite{zou2022devil} & $131.0_{\pm 1.6}$ & $188.4_{\pm 2.6}$ & $349.19$ & $296.15$\\
Our & $154.1_{\pm 1.7}$ & $206.8_{\pm 2.5}$ & $348.38$ & $296.07$\\
\bottomrule
\end{tabular}
}
\caption{
Comparison of the averaged encoding and decoding time, FLOPs and Memory footprint on Kodak dataset using a GPU (NVIDIA A40).
}
\label{tab:compl}
\end{table}

\subsection{Rate-Distortion Comparison}
\label{sec:rdcomparison}

Fig.~\ref{fig:rd_res} shows the rate-distortion curves for both Kodak and CLIC datasets. 
Our method improves over Zou2022~\cite{zou2022devil}, where the local attention maps are computed differently.
Regarding this architecture, \method~yields a BD-Rate gain of about $1.50$\% on Kodak and $0.89$\% on CLIC in terms of Bjontegaard metrics.
In particular, we perform especially better in the top-right part of the rate-distortion curves, at high PSNR and bit rate regime. 
Our analysis showed that our attention mechanism indeed tries to capture irregular local shapes trying to preserve them.
Thanks to this, fewer bits are allocated to high-contrast areas of the image, while preserving the correct level of details, but we struggle to represent flat regions.
Thus, for representing an image with high PSNR we required fewer bpps, but, since we use more bits to encode flat shapes, we cannot outperform in the high-compression regime, as we exemplify below.

\subsection{Allocation Maps Studies}

\begin{figure*}[!ht]
    \centering
    \begin{subfigure}{0.3\textwidth}
         \includegraphics[width=.95\linewidth]{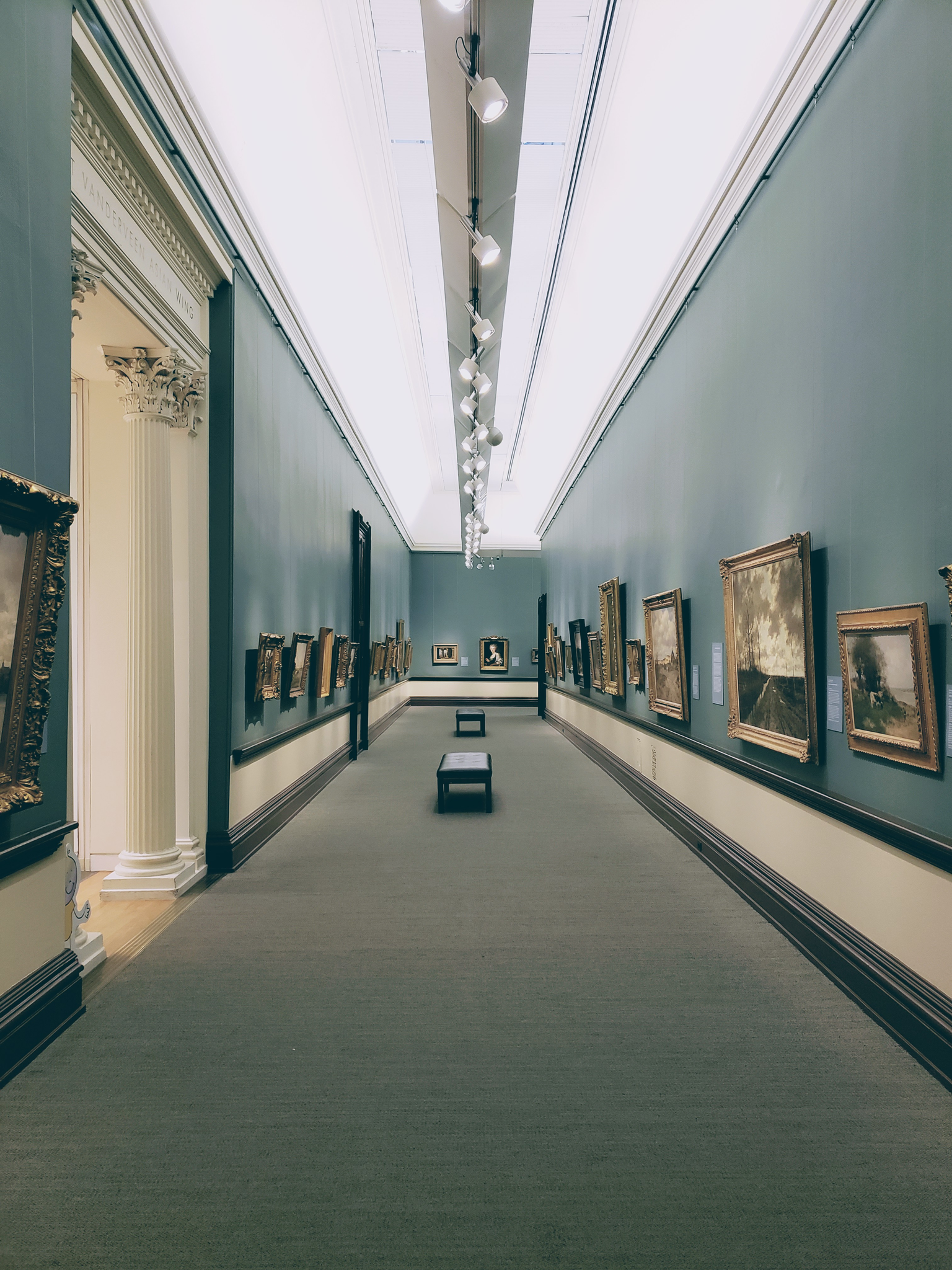}
         \caption{Original \texttt{image A}.}
     \end{subfigure}
     \hfill
    \begin{subfigure}{0.25\textwidth}
         \includegraphics[width=.9\linewidth]{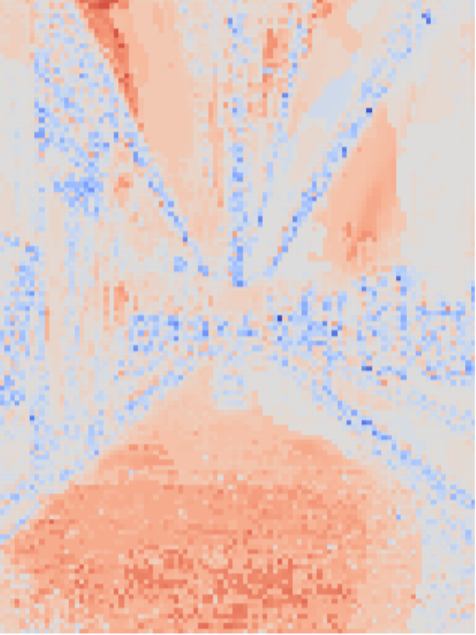}
         \caption{Bits allocation map for \texttt{A}\\
         Zou2022 0.48 bpp~~~PSNR:29.71\\
         \method~~~0.53 bpp~~~PSNR:29.88}
     \end{subfigure}
     \hfill
     \begin{subfigure}{0.25\textwidth}
         \includegraphics[width=.9\linewidth]{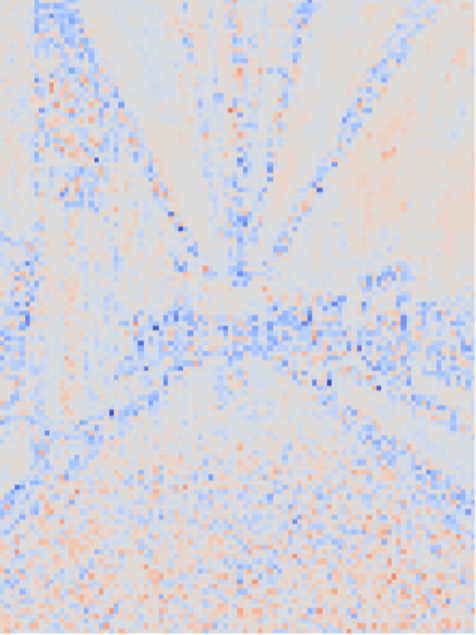}
         \caption{Bits allocation map for \texttt{A}\\
         Zou2022 1.21 bpp~~~PSNR:34.69\\
         \method~~~1.18 bpp~~~PSNR:34.99}
     \end{subfigure}
     \begin{subfigure}{0.3\textwidth}
         \includegraphics[width=.95\linewidth]{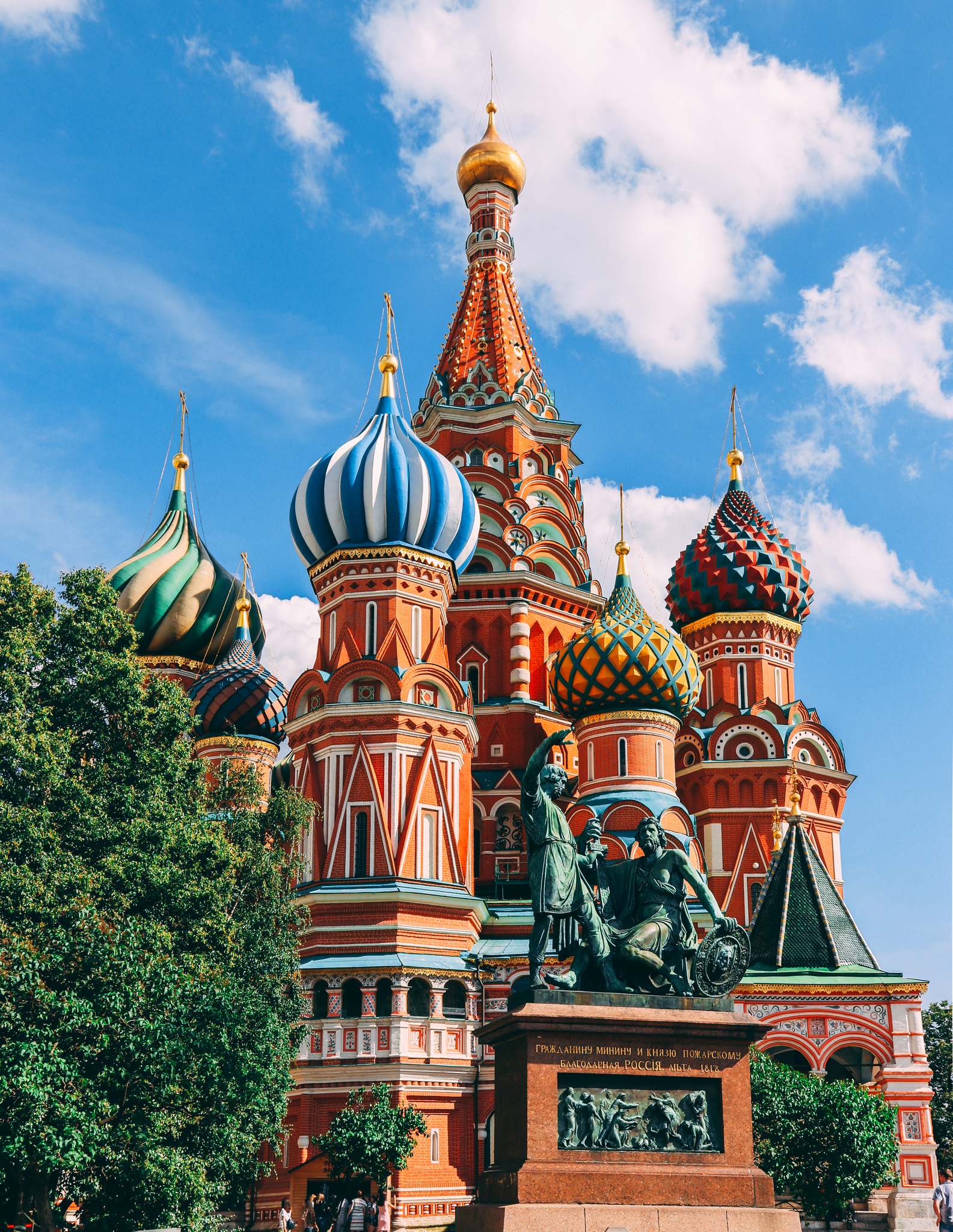}
         \caption{Original \texttt{image B}.}
     \end{subfigure}
     \hfill
    \begin{subfigure}{0.25\textwidth}
         \includegraphics[width=.9\linewidth]{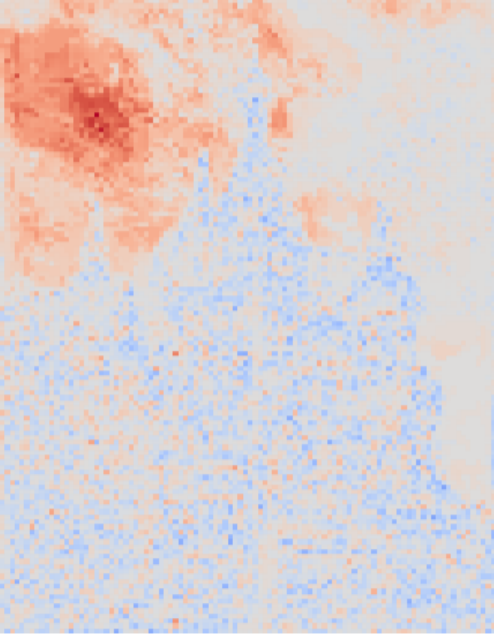}
         \caption{Bits allocation map for \texttt{B}\\
         Zou2022 0.10 bpp~~~PSNR:34.87\\
         \method~~~0.14 bpp~~~PSNR:34.89}
     \end{subfigure}
     \hfill
     \begin{subfigure}{0.25\textwidth}
         \includegraphics[width=.9\linewidth]{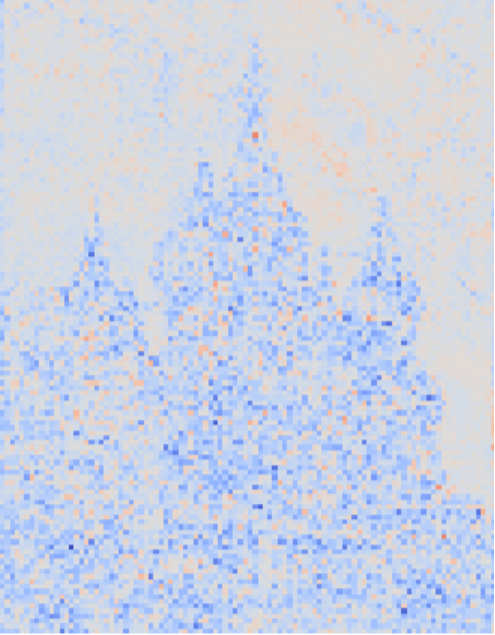}
         \caption{Bits allocation map for \texttt{B}\\
         Zou2022 0.40 bpp~~~PSNR:39.19\\
         \method~~~0.39 bpp~~~PSNR:39.26}
     \end{subfigure}
    \caption{Original images (a, d), bits allocation for low bitrates (b, e) and for high bitrates (c, f), where red indicates more bits for \method~and blue means less for \method~compared to Zou2022~\cite{zou2022devil}.}
    \label{fig:allocation_map}
\end{figure*}

Fig.~\ref{fig:allocation_map} shows the difference between the bit allocation map obtained using our Graph-based Window Block for coding concerning the use of a standard Window Block at different compression regimes. 
In blue are highlighted the areas of the image where \method~manages to allocate fewer bits than Zou2022~\cite{zou2022devil}, and in red the opposite.
We see that in a high-quality compression regime (top-right part of the RD curve), \method allocates fewer bits almost anywhere, especially in the high-frequency parts of the image that bear more perceptual importance and still obtain higher PSNR values.

However, at low-quality regimes the standard Window Block attention method is better at summarizing flat areas of the image, while our method seems to be trying to capture details in these areas as well, allocating more bits than necessary for this compression regime. Indeed, in this case, we also obtained higher PSNR values paid with a slightly higher bitrate.

\section{Conclusions}
\label{sec:conclusions}

In this paper, we have studied how local attention mechanisms work in image compression, and we have introduced \method, a novel methodology to compute the attention based on a graph constructed in each local window independently.
Through extensive evaluations on well-established datasets like Kodak and CLIC, we demonstrated that \method~achieves slightly superior results compared to state-of-the-art models.
Our gains come mainly at the more challenging high bit rates, where the fidelity is already high.
Our method excels in capturing irregular local shapes and preserving them, which is reflected in the improved performance.
To further highlight the capabilities of \method, we conducted a detailed study comparing the allocation bit maps produced by our model with those from models employing standard local attention mechanisms, with empirical results supporting our hypotheses.
However, \method~exhibits limitations at low bit rates, where the bit allocation fails to prioritize low-frequency features effectively.
Our future research endeavors will focus on addressing this challenge by incorporating a mechanism capable of capturing regular and lower-frequency shapes simultaneously. 
This approach aims to strike the optimal balance between preserving local details and prioritizing essential low-frequency components, thereby enhancing overall compression performance.

\section{Acknowledgements} 
This research was partially funded by Hi!PARIS Center on Data Analytics and Artificial Intelligence. This project was provided with computer and storage resources by GENCI at IDRIS thanks to the grant 2024-AD011015338 on the supercomputer Jean Zay's the V100 partition.

\bibliographystyle{IEEEbib}
\bibliography{main}

\end{document}